\documentclass[prl,aps,preprintnumbers,showpacs,superscriptaddress]{revtex4}
\usepackage{amsfonts,amssymb,amsmath}            % for math symbols.
\usepackage{graphics,graphicx,epsfig}            % for graphics figures.
\usepackage{amsthm}                              % for theorem environments

\def\openone{\leavevmode\hbox{\small1\normalsize\kern-.33em1}}

\newcommand{\tr}[1]{\mbox{Tr} \, #1 }
\newcommand{\be}{\begin{equation}}
\newcommand{\ee}{\end{equation}}
\newcommand{\bea}{\begin{eqnarray}}
\newcommand{\eea}{\end{eqnarray}}

\newcommand \s {\widetilde{s}}

\def\reff#1{(\ref{#1})}
\def\bbbc{{\mathchoice {\setbox0=\hbox{$\displaystyle\rm C$}\hbox{\hbox
to0pt{\kern0.4\wd0\vrule height0.9\ht0\hss}\box0}}
{\setbox0=\hbox{$\textstyle\rm C$}\hbox{\hbox
to0pt{\kern0.4\wd0\vrule height0.9\ht0\hss}\box0}}
{\setbox0=\hbox{$\scriptstyle\rm C$}\hbox{\hbox
to0pt{\kern0.4\wd0\vrule height0.9\ht0\hss}\box0}}
{\setbox0=\hbox{$\scriptscriptstyle\rm C$}\hbox{\hbox
to0pt{\kern0.4\wd0\vrule height0.9\ht0\hss}\box0}}}}
\newtheorem{theorem}{Theorem}

\def\bbbc{{\rm I\!C}}

\def \iden {\bf{I}}

\begin{document}
% Title & Authors

\title{Additivity for transpose depolarizing channels}

%\date{\today}
%\date[]{September 22, 2003}
%\date[\bf DRAFT:\rm\ ]{1-Jan-04}

\author{Nilanjana \surname{Datta}}
\email[]{n.datta@statslab.cam.ac.uk}
\affiliation{Statistical Laboratory,
             Centre for Mathematical Science,
             University of Cambridge,
             Wilberforce Road,
             Cambridge CB3 0WB, UK }

\author{Alexander S. \surname{Holevo}}
\email[]{holevo@mi.ras.ru}
\affiliation{Steklov Mathematical Institute,
             Gubkina 8,
             119991 Moscow,
             Russia}

\author{Yuri \surname{Suhov}}
\email[]{yms@statslab.cam.ac.uk}
\affiliation{Statistical Laboratory,
             Centre for Mathematical Science,
             University of Cambridge,
             Wilberforce Road,
             Cambridge CB3 0WB, UK }

%\date{July 11, 2002}

\begin{abstract}
Additivity of the minimal output entropy for the family of transpose
depolarizing channels introduced by Fannes et al. \cite{fannes} is
considered. It is shown that using the method of our previous paper
\cite{dhs} allows us to prove the additivity for the range of the
parameter values for which the problem was left open in
\cite{fannes}. Together with the result of \cite{fannes}, this
covers the whole family of transpose depolarizing channels.
\end{abstract}


\maketitle

\bigskip

\section{Introduction}

In a recent paper \cite{fannes} Fannes et al. considered the one
parameter family of \emph{transpose depolarizing channels}
 \be
\Phi(\mu) = t \mu^T + (1-t) \tr \mu \frac{{\iden}}{d},
\label{channel} \ee where \be - \frac{1}{d - 1} \le t \le
\frac{1}{d+1}. \label{range2} \ee Here $\mu$ is an arbitrary complex $d
\times d$ matrix, $\mu^{T}$ denotes its transpose, and ${\iden}$ is
the $d\times d$ unit matrix. The channel $\Phi $ is irreducibly
covariant since for any arbitrary unitary transformation $U$
\begin{equation}
\Phi (U\mu U^{\ast })=\bar{U}\Phi (\mu )\bar{U}^{\ast }, \label{cov}
\end{equation}
where $\bar{U}$ is the complex--conjugate of $U$ in a fixed basis.
Note that $\Phi(\mu)$ can be written as \be \Phi(\mu) = c\left(t +
\frac{1}{d-1}\right) \Phi_+(\mu) - c\left(t - \frac{1}{d+1}\right)
\Phi_-(\mu), \label{decomp} \ee where $c= (d^2 - 1)/2d$ and \be
\Phi_\pm (\mu):= \frac{1}{d\pm 1}\left({\iden} \tr \mu \pm
\mu^T\right) \ee The channels $\Phi_\pm (\mu)$ admit the following
Kraus decompositions \be \Phi_\pm (\mu) = \frac{1}{2(d\pm 1)}
\sum_{i,j=1}^d \left( |i\rangle\langle j| \pm |j\rangle\langle
i|\right) \mu \left( |i\rangle\langle j| \pm |j\rangle\langle
i|\right)^*. \ee From \reff{decomp} it follows that the channel
$\Phi$ interpolates between the channels $\Phi_+$ and $\Phi_-$,
where $\Phi_-$ is the Werner--Holevo channel introduced in \cite{HW}
and studied extensively (see e.g. \cite{my, dhs, af}).

Fannes et al. proved additivity of the minimal output entropy of the
channels (\ref{channel}) for \be - \frac{2}{d^2 - 2} \le t \le
\frac{1}{d+1}. \label{range1} \ee The values of the parameter $t$
given by \reff{range1} does not however cover the full range of
values (\ref{range2}). The aim of this paper is to extend the
validity of the additivity relation for the whole range of values of
$t$ given by \reff{range2}. More precisely, we prove additivity of
the minimum output entropy for \be - \frac{1}{d - 1} \le t \le 0.
\label{range3} \ee

The minimum output entropy of a channel $\Phi$ is
\begin{equation}
h(\Phi ):=\min_{\rho } S(\Phi (\rho )), \label{minent1}
\end{equation}
where the minimization is over all possible input states $\rho$
(i.e., density matrices) of the channel. Here and below $S(\sigma )
:=-\mbox{Tr}\sigma \log \sigma $ denotes the von Neumann entropy of
the density matrix $\sigma $. Fannes et al. proved the additivity
relation \be h(\Phi \otimes \Phi ) = 2h(\Phi) \label{addd} \ee for
the values of $t$ given by \reff{range1}.   For simplicity of
exposition we also consider the case $d_1=d_2 = d$ although the
proof can
 be easily extended to the case $d_1 \ne d_2$, i. e. \be h(\Phi_1 \otimes \Phi_2 ) =
h(\Phi_1) + h(\Phi_2). \label{addrel} \ee The proof employs the
method developed in \cite{dhs}.

Consider the Schmidt decomposition
\begin{equation}
|\ \psi _{12}\rangle =\sum_{\alpha =1}^{d}\sqrt{\lambda _{\alpha
}}|\alpha ;1\rangle \otimes|\alpha ;2\rangle .  \label{schmidt}
\end{equation}
Here $\left\{ |\alpha ;j\rangle \right\} $ is an orthonormal basis in ${\
\mathcal{H}}_{j}$, $j=1,2$, and ${\underline{\lambda }}=(\lambda _{1},\ldots
,\lambda _{d})$ is the vector of the Schmidt coefficients. The state $|\psi
_{12}\rangle \langle \psi _{12}|$ can then be expressed as
\begin{equation}
\label{schmidt2}
|\psi _{12}\rangle \langle \psi _{12}|=\sum_{\alpha ,\beta =1}^{d}\sqrt{
\lambda _{\alpha }\lambda _{\beta }}|\alpha ;1\rangle \langle \beta
;1|\otimes |\alpha ;2\rangle \langle \beta ;2|.
\end{equation}
The Schmidt coefficients form a probability distribution:
\begin{equation}
\lambda _{\alpha }\geq 0\quad ;\quad \sum_{\alpha =1}^{d}\lambda _{\alpha
}=1;  \label{nine}
\end{equation}
thus the vector ${\underline{\lambda }}$ varies in the $({d}-1)-$dimensional
simplex $\Sigma _{d}$, defined by these constraints. The extreme points
(vertices) of $\Sigma _{d}$ correspond precisely to unentangled vectors $
|\psi _{12}\rangle =|\psi _{1}\rangle \otimes |\psi _{2}\rangle \in {
\mathcal{H}}_{1}\otimes {\mathcal{H}}_{2}$. Then the additivity (\ref{addd})
follows
if for every choice of the bases $
\left\{ |\alpha ;1\rangle \right\} $ and $\left\{ |\alpha ;2\rangle \right\}
$, the function
\begin{equation}
\underline{\lambda }\rightarrow S\left( \sigma_{12}({\ \underline{
\lambda }})\right),  \label{map}
\end{equation}
where
\begin{equation}
\sigma_{12}({\underline{\lambda }}):=\left( \Phi \otimes \Phi \right)
\left( |\psi _{12}\rangle \langle \psi _{12}|\right) =\sum_{\alpha ,\beta
=1}^{d}\sqrt{\lambda _{\alpha }\lambda _{\beta }}\Phi (|\alpha ;1\rangle
\langle \beta ;1|)\otimes \Phi (|\alpha ;2\rangle \langle \beta ;2|).
\label{matrix}
\end{equation}
is the channel
output state, attains its minimum at the vertices of $\Sigma _{d}$.
Owing to
(\ref{cov}), we can choose for $\{|\alpha ;i\rangle\}$
the canonical basis of real vectors $\left\{ |\alpha
\rangle \right\} $ in ${\mathcal{H}}_{i}\simeq {\mathbf{C}}^{d_{i}};$ $i=1,2$.
Moreover from the definition \reff{channel} of the channel $\Phi$ it follows
that
\begin{equation*}
\Phi \left( |\alpha \rangle \langle \beta |\right) =
(1-t) \delta _{\alpha \beta }\frac{{\iden}}{d} + t |\beta \rangle \langle \alpha |,
\end{equation*}
since $|\alpha \rangle $ and $|\beta \rangle $ are real.
Hence,
\bea
\sigma_{12}({\underline{\lambda }})&=& \sum_{\alpha ,\beta
=1}^{d}\sqrt{\lambda _{\alpha }\lambda _{\beta }}\Phi(|\alpha \rangle
\langle \beta |)\otimes \Phi (|\alpha \rangle \langle \beta |)\nonumber\\
&=& \sum_{\alpha ,\beta =1}^{d}|\alpha \beta \rangle
\langle \alpha \beta |\left[ \frac{(1-t)^2}{d^2} + \frac{t(1-t)}{d}
\left(\lambda _{\alpha }+ \lambda _{\beta }\right)
\right] +\sum_{\alpha
,\beta =1}^{d} t^2 \,\sqrt{\lambda _{\alpha }\lambda _{\beta }}|\alpha \alpha
\rangle \langle \beta \beta |.  \label{matrixn}
\eea
Here we have used the constraint \reff{nine} and the fact that
${\iden}=\sum_{\alpha =1}^{d}|\alpha \rangle \langle \alpha |$.

To find the minimum output entropy of the product channel $\Phi
\otimes \Phi$, we first evaluate the eigenvalues of
$\sigma_{12}({\underline{\lambda }})$. For this purpose it is useful
to express $\sigma_{12}({\underline{\lambda }})$ in the form of a
$d^2 \times d^2$ matrix $A$ with elements \be A_{ij} = (\mu_i +
\eta_i)\delta_{ij} + \sqrt{\eta_i \eta_j} (1 - \delta_{ij}), \ee
where we identify $i$ or $j$ with a pair $(\alpha ,\beta )$ and
define
\begin{equation}
\mu _{i}\equiv \mu _{\alpha \beta }=\frac{(1-t)^2}{d^2} + \frac{t(1-t)}{d}
\left(\lambda _{\alpha }+ \lambda _{\beta }\right)
\quad ;\quad \eta _{j}\equiv \eta _{\alpha \beta }=\lambda _{\alpha } t^2 \delta
_{\alpha \beta },\quad \alpha ,\beta =1,\ldots ,{d}.  \label{parameters}
\end{equation}
As shown in \cite{dhs}, the characteristic equation
${\hbox{det}}(A-\gamma {\iden})=0$ can be written as
\be \prod_{1
\le \alpha, \beta\le d\atop{\alpha \ne \beta}}(\mu
_{\alpha\beta}-\gamma ) \left[ \prod_{\alpha ^{\prime }=1}^{d}(\mu
_{\alpha'\alpha'}-\gamma )\left\{ 1+\sum_{\alpha ^{^{\prime \prime
}}=1}^{d}\frac{t^2 \lambda _{\alpha''}}{(\mu_{\alpha''\alpha''} -\gamma )}\right\}
\right] = 0.  \notag \\
\ee
This implies that $\sigma_{12}({\underline{\lambda }})$ has the following sets of
eigenvalues:
\begin{enumerate}
\item{ $d(d-1)$ eigenvalues of the form
\be
\gamma_{\alpha\beta} = \mu_{\alpha\beta} =  \frac{(1-t)^2}{d^2} + \frac{t(1-t)}{d}\left(\lambda _{\alpha }+ \lambda _{\beta }\right),  \quad \alpha \ne \beta, \, \alpha,\beta= 1,\ldots, d.
\label{eig1}
\ee}
\item{$d$ eigenvalues $\{g _{\alpha }, \alpha =1,\dots ,d\}$,
given by the roots of the equation
\be
\prod_{\alpha
=1}^{d}(\mu _{\alpha\alpha}-g )\left\{
1+\sum_{\alpha'=1}^{d}\frac{t^2 \lambda _{\alpha'}}
{(\mu_{\alpha'\alpha'} -g )}\right\}
= 0.
\ee
This equation can be written as
\begin{equation}
\prod_{\alpha =1}^{d}(c_1 + c_2 \lambda _{\alpha }- g )\left\{ 1+\sum_{\alpha
^{\prime }=1}^{d}\frac{ t^2 \lambda _{\alpha ^{\prime }}}{(c_1 + c_2\lambda _{\alpha
^{\prime }}- g )}\right\} =0  \label{eigen2}.
\end{equation}}
\end{enumerate}
Here we have defined \be c_1 = \frac{(1-t)^2}{d^2}  \quad ; \quad
c_2 = \frac{2t(1-t)}{d}. \label{c1c2} \ee Since $t$ is in the range
\reff{range3},
\begin{equation}\label{range4}
    c_2 \le 0,\quad -2 \le c_2/c_1=\frac{2td}{1-t} \le 0.
\end{equation}

The von Neumann entropy of the output of the product channel can be
expressed as a sum
\begin{equation}
S(\sigma_{12}({\underline{\lambda }}))=S_{1}({\underline{\lambda }})+S_{2}({
\underline{\lambda }})  \label{sum2}
\end{equation}
where
\begin{equation}
S_{1}({\underline{\lambda }}):=-\sum_{1\leq \alpha ,\beta \leq d\atop{
\, \beta \neq \alpha }}\gamma_{\alpha \beta }\log \gamma_{\alpha \beta },\;\;\quad S_{2}(
{\underline{\lambda }}):=-\sum_{\alpha =1}^{d}g_{\alpha }\log g_{\alpha
}.  \label{s11}
\end{equation}
Note that \be \sum_{1\leq \alpha ,\beta \leq d\atop{\beta \neq
\alpha}}\gamma_{\alpha \beta } = \frac{d-1}{d} (1-t)^2:= c. \ee
Moreover, using the fact that the eigenvalues of
$\sigma_{12}({\underline{\lambda }})$ sum to $1$ we get \be
\sum_{\alpha=1}^d g_\alpha = 1 - c. \ee

Using the above relations we can define sets of non--negative
variables \be {\widetilde{\gamma_{\alpha \beta }}} :=
\frac{1}{c}\gamma_{\alpha \beta },\, \alpha \ne \beta, \quad \alpha,
\beta = 1, \ldots, d. \label{wide1} \ee and \be
{\widetilde{g_{\alpha}}} := \frac{1}{1-c}g_{\alpha} \quad ; \quad
\alpha=1.2,\ldots d, \label{wide2} \ee each of which sum to unity,
i.e.,
$$\sum_{1\leq \alpha ,\beta \leq d \atop{\beta \neq \alpha }}
{\widetilde{\gamma_{\alpha\beta}}} = 1 \quad; \quad
\sum_{\alpha=1}^d {\widetilde{g_{\alpha}}} =1, $$ and hence define
probability distributions. In terms of these variables we have \bea
S_{1}({\underline{\lambda }}) &=& c
H(\{{\widetilde{\gamma_{\alpha \beta }}} \}) + {\hbox{\, const}}\label{s110}\\
S_{2}({\underline{\lambda }}) &=& (1-c)
H(\{{\widetilde{g_{\alpha}}} \}) + {\hbox{\, const}}\label{s22}
\eea
Here $H(\{{x_i}\})$ denotes the Shannon entropy of a probability distribution
$\{x_1, \ldots, x_n : x_i\ge 0, \sum_{i=1}^n x_i =1\}$. Since $H(\{{x_i}\})$ is a concave function of the variables
$x_i, i=1,\ldots, n$, it follows from \reff{s110} that $S_{1}({\underline{\lambda }})$
is a symmetric concave function of the variables ${\widetilde{\gamma_{\alpha \beta }}}$.
These variables (defined by \reff{wide1} and \reff{eig1}) are affine functions of the Schmidt
coefficients $\lambda _{1},\ldots ,\lambda _{d}$. Hence, $S_{1}$ is a
concave function of $\ {\underline{\lambda }}$ and attains its global
minimum at the vertices of the simplex $\Sigma _{d}$, defined by the
constraints (\ref{nine}).

Let us now analyze $S_{2}$. We wish to prove the following:

\begin{theorem}
The function $S_{2}$ is Schur-concave in $\underline{ \lambda }\in
\Sigma_{d} $ i.e., $\underline{\lambda }\prec \underline{\lambda
}^{\prime }\, \implies S_{2}\left( {\ \underline{\lambda }}\right)
\geq S_{2}\left( {\underline{ \lambda }}^{\prime }\right)$, where
$\prec $ {denotes the stochastic majorization} (see \cite{bha}).
\end{theorem}

Since every $\underline{\lambda }\in \Sigma _{d}$ is majorized by
the vertices of $\Sigma _{d}$, this will imply that
$S_{2}(\underline{\lambda })$ also attains its minimum at the
vertices. Thus $S(\underline{\lambda } )=S_{1}(\underline{\lambda
})+S_{2}(\underline{\lambda })$ is minimized at the vertices, which
correspond to unentangled states. As was observed, this implies the
additivity (\ref{addd}).

%%%%%%%%%%%%%%%%
\section{Proof of the Theorem}

In \cite{graeme} it was proved that the
Shannon entropy $H({\underline{x}})$, $\underline{x}=(x_{1},\ldots
,x_{d})\in \Sigma _{d}$, is a monotonically non--decreasing function of the
elementary symmetric polynomials $s_{q}(x_{1},x_{2},\ldots ,x_{d})$ (see
e.g. \cite{bha}) in the variables $x_{1},x_{2}$, $\ldots $, $x_{d}$, $q=2,\ldots ,d$.
This implies that $S_{2}$ is a monotonically non--decreasing function of the
symmetric polynomials
\begin{equation}
\s_{q}({\underline{\lambda }}):=s_{q}(g_{1},g_{2},\ldots ,g_{d}),\quad q=2,\ldots ,d.  \label{wides}
\end{equation}
Therefore, to prove the Theorem it is sufficient to prove that the
functions $\s_{q}({\underline{\lambda }})$ are Schur concave in
$\underline{\lambda }\in \Sigma _{d}$ for $q=2, \ldots, d$.

Let us define the variables
\begin{equation}
\nu _{\alpha }:= 1 + \frac{c_2}{c_1}\lambda _{\alpha },\quad \alpha =1,2,\ldots ,d.
\label{nu}
\end{equation}
This together with \reff{nine} implies that \be \label{cons2}
1+c_2/c_1\le\nu_\alpha \le 1,\qquad \sum_{\alpha = 1}^d \nu_\alpha
=d+ c_2/c_1. \ee Defining $\gamma = g/c_1$, (\ref{eigen2}) can be
expressed in terms of the variables $\nu _{\alpha }$ as follows
\begin{equation}
\prod_{\alpha =1}^{d}(\nu _{\alpha }-\gamma )\left\{ 1+
\sum_{\alpha ^{\prime }=1}^{d}\frac{(\nu _{\alpha ^{\prime }}-1)t^2}{c_2 (\nu
_{\alpha ^{\prime }}-\gamma )}\right\} =0.  \label{twonu}
\end{equation}

Denote $\gamma_i := g_i/c_1$ for $i=1,\ldots, n$, where $g_1, g_2,
\ldots, g_n$ are the roots of eq.\reff{eigen2}. Therefore $\gamma
_{1},\ldots ,\gamma _{d}$ are the zeroes of the product $(\gamma
_{1}-\gamma )(\gamma _{2}-\gamma )\ldots (\gamma _{d}-\gamma )$ and
equation (\ref{twonu}) can be expressed in terms of these roots as
follows:
\begin{equation}
\sum_{k=0}^{d}\gamma ^{k}\,(-1)^{k}\,s_{d-k}(\gamma _{1},\gamma _{2},\ldots
,\gamma _{d})=0.  \label{sym1}
\end{equation}
In terms of the elementary symmetric polynomials $s_{l}$ of the
variables $ \nu _{1},\nu _{2},\ldots ,\nu _{d}$, (\ref{twonu}) can
be rewritten as
\begin{equation}
\sum_{k=0}^{d}\gamma ^{k}\,(-1)^{k}\,s_{d-k}(\nu _{1},\nu _{2},\ldots ,\nu
_{d})+\sum_{k=0}^{d-1}\gamma ^{k}\,(-1)^{k}\,\sum_{l=1}^{d}s_{d-1-k}(\nu
_{1},\ldots ,{\not{\nu _{l}}}\ldots ,\nu _{d})\,\frac{(\nu _{l}-1)t^2}{c_2}=0,
\label{sym3}
\end{equation}
where the symbol ${\not{\nu _{l}}}$ means that the variable ${\nu _{l}}$ has
been omitted from the arguments of the corresponding polynomial. Equating
the LHS of (\ref{sym1}) with the LHS of (\ref{sym3}) yields, for each $0\leq
k\leq d-1$ :
\begin{equation}
s_{d-k}(\gamma _{1},\gamma _{2},\ldots ,\gamma _{d})=s_{d-k}(\nu _{1},\nu
_{2},\ldots ,\nu _{d})+\sum_{l=1}^{d}s_{d-1-k}(\nu _{1},\ldots ,{\not{\nu
_{l}}}\ldots ,\nu _{d})\,\frac{(\nu _{l}-1)t^2}{c_2}.  \label{sympol}
\end{equation}
Note that in (\ref{sympol}), values $s_{d-k}(\gamma _{1},\gamma _{2},\ldots
,\gamma _{d})$ are expressed in terms of values of elementary symmetric
polynomials in the variables $\nu _{1},\nu _{2},\ldots ,\nu _{d}$ (which are
themselves linear functions of the Schmidt coefficients $\lambda _{1},\ldots
,\lambda _{d}$).

Our aim is to prove that $\s_{q}(\underline{\lambda })$ is Schur concave in
the Schmidt coefficients $\lambda_{1},\ldots ,\lambda_{d}$,
for $q=2, \ldots, d$. Eq.(\ref{wides})
implies that this amounts to proving Schur concavity of $s_{d-k}(\gamma_{1},
\gamma_{2},\ldots , \gamma_{d})$ as a function of $\lambda_{1},\ldots
,\lambda_{d}$, for all $0 \le k \le d-2$. The functions
\begin{equation}
\Phi_{k}(\nu_{1},\ldots ,\nu_{d}):=s_{d-k}(\nu_{1},\ldots ,\nu
_{d})+\sum_{l=1}^{d}s_{d-1-k}(\nu_{1},\ldots ,{\not{\nu_{l}}}\ldots ,\nu
_{d})\,\frac{(\nu _{l}-1)t^2}{c_2}\equiv {\hbox{RHS of }}(\ref{sympol})
\label{phidef}
\end{equation}
are symmetric in the variables $\nu_{1},\nu_{2},\ldots ,\nu_{d}$,
and hence in the variables $\lambda_{1},\ldots ,\lambda_{d}$. By the
necessary and sufficient condition for Schur concavity \cite{bha} it
is enough to prove
\begin{equation}
(\lambda_{i}-\lambda _{j})\bigl(\frac{\partial \Phi_{k}}{\partial\lambda_{i}}
-\frac{\partial \Phi_{k}}{\partial\lambda_{j}}\bigr) \equiv (\nu
_{i}-\nu_{j})\bigl(\frac{\partial \Phi_{k}}{\partial \nu_{i} }-\frac{
\partial \Phi_{k}}{\partial \nu_{j}}\bigr)\leq 0,\quad \forall \,1\leq
i,j\leq d.  \label{key2}
\end{equation}

By the rule of differentiation of the elementary symmetric
polynomials, see e.g. \cite{bha}, we have
\begin{eqnarray}
\frac{\partial }{\partial \nu_{i}}\Phi_{k}(\nu _{1},\ldots ,\nu_{d})&=&
\frac{\partial }{\partial \nu_{i}}s_{d-k}(\nu _{1},\ldots ,\nu_{d})+\frac{
\partial }{\partial \nu _{i}}\sum_{l=1}^{d}s_{d-1-k}(\nu_{1},\ldots ,{\not{
\nu_{l}}}\ldots ,\nu _{d})\,\frac{(\nu _{l}-1)t^2}{c_2}  \notag \\
&=&s_{d-1-k}(\nu _{1},..,{\not{\nu_{i}}},..,\nu_{d})+
\sum_{1\leq
l\leq d \atop{l\neq i}}s_{{d-1-k}}(\nu_{1},\ldots ,{\not{\nu_{i}}},..,{\not{
\nu_{l}}} \ldots ,\nu_{d})\,\frac{(\nu _{l}-1)t^2}{c_2}  \notag \\
&&+\frac{t^2}{2}s_{d-1-k}(\nu_{1},\ldots ,{\not{\nu_{i}}}\ldots ,\nu _{d}).
\label{eq1}
\end{eqnarray}
Therefore,
\begin{eqnarray}
\Bigl(\frac{\partial \Phi_{k}}{\partial \nu_{i}}&-&\frac{\partial \Phi_{k}}{
\partial \nu_{j}}\Bigr)(\nu_{1},\ldots ,\nu_{d}) =s_{d-1-k}(\nu_{1},.., {
\not{\nu_{i}}},..,\nu_{d})-s_{d-1-k}(\nu_{1},..,{\not{\nu_{j}}},..,\nu _{d})
\notag \\
&+&\sum_{1\leq l\leq d\atop{l\neq i}}s_{d-1-k}(\nu_{1},..,{\not{
\nu_{i}}} ,..,{\not{\nu_{l}}}..,\nu_{d})\,\frac{(\nu_{l} - 1)t^2}{c_2} -
\sum_{1\leq l\leq d\atop{ l\neq j}}s_{d-1-k}(\nu_{1},..,{\not{\nu_{j}}},..,{
\not{\nu_{l}}} \ldots ,\nu_{d})\,\frac{(\nu_{l} - 1)t^2}{c_2}  \notag \\
&+&\frac{t^2}{c_2}\bigl[s_{d-1-k}(\nu_{1},\ldots ,{\not{\nu_{i}}}\ldots ,\nu
_{d})-s_{d-1-k}(\nu_{1},\ldots ,{\not{\nu_{j}}}\ldots ,\nu_{d})\bigr].
\end{eqnarray}
Using a transformation rule for the elementary symmetric
polynomials, see e.g. \cite{bha}, we get
\begin{eqnarray}
\Bigl(\frac{\partial \Phi_{k}}{\partial \nu_{i}}&-&\frac{\partial \Phi_{k}}{
\partial \nu_{j}}\Bigr)(\nu_{1},\ldots ,\nu_{d}) = \bigl(\nu
_{j}-\nu_{i})s_{d-k-2}(\nu_{1},..,{\not{\nu_{i}}},{\not{\nu_{j}}} \ldots
,\nu_{d})  \notag \\
&+&\frac{2t^2}{c_2}{(\nu_{j}-\nu_{i})}s_{d-k-2}(\nu_{1},..,
{\not{\nu_{i}}},{\not{\nu_{j}}}\ldots ,\nu_{d})  \notag \\
&+&\sum_{1\leq l\leq d\atop{l\neq i,j}}\frac{(\nu_{l} - 1)t^2}{c_2}\bigl[
s_{d-k-2}(\nu_{1},..,{\not{\nu_{i}}},..,{\not{\nu_{l}}}\ldots ,\nu
_{d})-s_{d-k-2}(\nu_{1},..,{\not{\nu_{j}}},..,{\not{\nu_{l}}}\ldots ,\nu
_{d})\bigr]  \notag \\
&=&\sum_{1\leq l\leq d\atop{l\neq i,j}}\frac{(\nu_{l} - 1)t^2}{c_2}
(\nu_{j}-\nu _{i})s_{d-k-3}(\nu_{1},..,{\not{\nu_{i}}},..,{\not{\nu_{j}}},..{
\not{\nu _{l}}}\ldots ,\nu_{d})\nonumber\\
&+& \bigl(1 + \frac{2t^2}{c_2}\bigr){(\nu_{j}-\nu_{i})}s_{d-k-2}(\nu_{1},..,
{\not{\nu_{i}}},..,{\not{\nu_{j}}}\ldots ,\nu_{d}).  \label{long22}
\end{eqnarray}
Substituting (\ref{long22}) in (\ref{key2}), using (\ref{c1c2}) and
rearranging factors, we obtain that the Schur concavity holds if and
only if \be \sum_{1\leq l\leq d\atop{l \neq i,j}} (1-\nu_{l} )
s_{d-k-3}(\nu_{1},..,{
\not{\nu_{i}}},{\not{\nu_{j}}},..{\not{\nu_{l}}}\ldots ,\nu_{d})
-\frac{2(1+t(d-1))}{td}
s_{d-k-2}(\nu_{1},..,{\not{\nu_{i}}},{\not{\nu_{j}}} \ldots
,\nu_{d})\geq 0, \label{main} \ee for all $1\leq i,j\leq d$ and
$0\le k \le d-2$. The
variables $\nu_{i}$ and $\nu_{j}$ do not appear in (\ref{main}).
Owing to symmetry, without loss of generality, we can choose $i=d-1$
and $j=d$. Then omitting $\nu_{d-1}$ and $\nu_{d}$ and  setting
$n=d-2$, we obtain that the functions $\Phi_{k}$ defined in
(\ref{phidef}) are Schur concave in the Schmidt coefficients
$\lambda_1, \ldots,\lambda_d$ if and only if
\begin{equation}
\sum_{l=1}^{n}(1-\nu_{l})s_{n-k-1}(\nu_{1},..,{\not{\nu_{l}}}\ldots
,\nu _{n})-\frac{2(1+t(d-1))}{td}s_{n-k}(\nu_{1},..,\nu_{n})\geq 0,
\label{main2}
\end{equation}
for $0 \le k \le n-1$. Here the variables $\nu_{l}$, $1\le l \le n$,
satisfy the constraints \be  \nu_l \le 1 ;\quad
\sum_{l=1}^{n}\nu_{l} \geq n+c_2/c_1=n+\frac{2td}{1-t}\ge n-2,
\label{1nu1} \ee following from (\ref{nu}), the relations:\,
$\lambda_l \ge 0$ for all $l$, and $
\displaystyle{\sum_{l=1}^{n}\lambda_{l}=\sum_{l=1}^{d-2}\lambda_{l}\leq
1}$, and \reff{range4}.

The above constraint implies that $1-\nu_{l} \le 0$ for $1\leq l \leq n$.
Thus if {all $\nu _{1},\ldots ,\nu_{n}\geq 0,$ (\ref{main2})
obviously holds. The constraint \reff{1nu1} also implies that
\emph{at most one} of the variables $\nu_{1},\ldots ,\nu_{n}$ can be
{\em{negative}}. Hence, we need to prove (\ref{main2}) only in the
case in which one and only one of the variables $ \nu_{1},\ldots
,\nu_{n}$ is negative.

We now proceed to prove \reff{main2}. We first notice that for all
values of $0\le k \le n$ nonnegativity of the \emph{ first term} in
the LHS of \reff{main2} under the constraints \be  \nu_l \le 1;
\quad \sum_{l=1}^{n}\nu_{l} \geq n-2, \label{2nu1} \ee which are
weaker than (\ref{1nu1}), and coincide with them for
${\displaystyle{t=-\frac{1}{d-1}}}$, was proven in \cite{dhs}. Next
we prove that the \emph{second term} on the LHS of \reff{main2} is
positive for $k=1,2, \ldots n$. These two facts together prove
\reff{main2} for all $k=1,2, \ldots,n$. For $k=0$ the second term is
not positive. In this case we prove \reff{main2} by considering the
sum of the two terms on the LHS of \reff{main2}.

Let us now analyze the second term on the LHS of \reff{main2} for $
1\le k \le n$. In the range (\ref{range3}) we have
\[-\frac{(1+t(d-1))}{td}\ge 0.\]
Also
\bea
&&s_{n-k}(\nu_{1},..,\nu_{n})\nonumber\\
&=&  \sum_{2 \le i_1<i_2<...<i_{n-k}\le n}
\nu_{i_1},..,\nu_{i_{n-k}} + \nu_1 \sum_{2 \le
i_1<i_2<...<i_{n-k-1}\le n}
\nu_{i_1},..,\nu_{i_{n-k-1}}\nonumber\\
&=&
\frac{1}{(n-k-1)!}\sum_{i_1,i_2,..i_{n-k-1}=2}^n\nu_{i_1},..,\nu_{i_{n-k-1}}
\bigl( \nu_1 + \sum_{r=2\atop{r \ne i_1, \ldots,i_{n-k-1}}}^n
\nu_r\bigr)
\nonumber\\
&=&  \frac{1}{(n-k-1)!}\sum_{i_1,i_2,..i_{n-k-1}=2}^n
\nu_{i_1},..,\nu_{i_{n-k-1}} \bigl(\nu_1 + \bigl\{\sum_{r=1}^n \nu_r
- \nu_1 - (\nu_{i_1}+..+\nu_{i_{n-k-1}}
\bigr\}\bigr)\nonumber\\
&\ge& \frac{1}{(n-k-1)!} \sum_{i_1,i_2,..i_{n-k-1}=2}^n
\nu_{i_1},..,\nu_{i_{n-k-1}} \times \bigl((n-2) - (n-k-1)\bigr)
\nonumber\\
&=&\frac{1}{(n-k-1)!} \sum_{i_1,i_2,..i_{n-k-1}=2}^n
\nu_{i_1},..,\nu_{i_{n-k-1}} \times \bigl(k-1\bigr) \ge 0,
\label{second} \eea since $k\geq 1$. Hence, \be \left[{\hbox{2nd
term on LHS of \reff{main2}}}\right] \ge 0 \quad {\hbox{for }} 1 \le
k \le n. \label{22} \ee In the second last line of eq.\reff{second}
we have used  the constraint  \reff{1nu1}. The negativity of the
first term on the LHS of \reff{main2} (as proved in \cite{dhs})
together with \reff{22} implies that the inequality \reff{main2}
holds for all
$k=1,2, \ldots, n$.

\noindent
{\em{ The case $k=0$}}:

In this case we have \be {\hbox{LHS of \reff{main2}}}= e_n
\left[\sum_{l=1}^n \frac{(1-\nu_l)}{ \nu_{l}}
-\frac{2(1+t(d-1))}{td} \right] \ee where $e_n:= \nu_1,..,{\nu_l}
\ldots, \nu_n <0$, since one and only one of the variables
$\nu_1, \ldots, \nu_n$ is negative. Hence in this case the inequality \reff{main2}
reduces to \be \sum_{l=1}^n \frac{(1-\nu_l)}{ \nu_{l}}
\le\frac{2(1+t(d-1))}{td}(\le 0). \label{last2} \ee

Without loss of generality we can choose $\nu_{1}<0$ and $\nu_{l}>
0$ for all $l=2,3,\ldots ,n$. The function
\begin{equation}\label{fnu}
 f(\nu)=\frac{1-\nu}{\nu}=\frac{1}{\nu}-1
\end{equation}
is nonincreasing for all $\nu$, convex for $\nu >0$ and $f(1)=0$.
Denote
\begin{equation}\label{gnu}
 g(\nu_2,\dots,\nu_n)=\sum_{l=2}^n f(\nu_l),
\end{equation}
then $g$ is convex on the simplex
\begin{equation}\label{simplex}
 \nu_2+\dots+\nu_n \ge n+c_2/c_1-\nu_1,\quad (0\le)\nu_l\le 1,
 \,\,l=2,\dots,n,
\end{equation}
where $\nu_1<0$ is fixed, and hence attains its maximum on its
extreme points. These are\begin{equation}\label{extremep}
 (2+c_2/c_1-\nu_1,1,\dots ,1)
\end{equation}
and its permutations, and $(1,1,\dots ,1)$. In the first case
\begin{equation}\label{nu12}
\nu_1+\nu_2=2+c_2/c_1=\frac{2(1+t(d-1))}{1-t},
\end{equation}
and we have to show that \begin{equation}\label{ziel}
 \frac{1}{\nu_1}+  \frac{1}{\nu_2}-2\le  \frac{2(1+t(d-1))}{td}.
\end{equation}
The second case reduces to this because it corresponds to $\nu_2=1$
(and $\nu_l=1,l>2$), and the LHS of \reff{ziel} is then maximal for
the minimal possible value $\nu_1=1+c_2/c_1$ (see (\ref{cons2})),
for which the condition (\ref{nu12}) is satisfied.

To prove (\ref{ziel}) we take into account that $\nu_1<0,\nu_2>0$.
Then it reduces to \begin{equation}\label{ziel2}
\frac{2(1+t(d-1))}{1-t}=\nu_1+\nu_2 \ge
2\nu_1\nu_2\left[\frac{1+t(d-1)}{td}+1\right].
\end{equation}
The product
$$\nu_1\nu_2=\nu_1\left[\frac{2(1+t(d-1))}{1-t}-\nu_1\right]$$ is
nonpositive and monotonically increases from the value
$$\nu_1^0=\frac{2(1+t(d-1))}{1-t}-1<0$$ to zero. Since the LHS of
(\ref{ziel2}) is nonnegative, it is sufficient, whatever
the sign of the last factor on the right hand side of (\ref{ziel2})
is, to check it only for $\nu_1=\nu_1^0$. Substituting this value
and making common denominator, we get, taking into account that
$t<0$,
\[(1+t(d-1))td\le 2(1+t(d-1))^2+(1+t(d-1))(2td-1+t))-(1-t)td\]
or \[0\le 2(1+t(d-1))^2+(td)^2 -(1-t)^2-(1-t)(td)\] or \[0\le
3(td)^2+3(1-t)(td)+(1-t)^2,\] which is indeed true.

\section{Acknowledgements}
The second author acknowledges partial support from the QIS Program of the
Newton Institute, and from A. von Humboldt Foundation.

\end{document}